\documentclass[11pt]{article}

\usepackage{amsmath,amssymb, amsthm}
\usepackage{mathrsfs}
\usepackage{framed}
\usepackage{lmodern}
\usepackage{todonotes}
\usepackage[ruled,vlined]{algorithm2e}
\usepackage{algpseudocode}
\usepackage{tikz}
\usepackage{tkz-graph}
\usepackage{pgfplots}
\usepackage{color}
\usepackage{authblk}
\usepackage{fullpage}

\numberwithin{equation}{section}
\newtheorem{theorem}{Theorem}
\numberwithin{theorem}{section}

\newtheorem{proposition}[theorem]{Proposition}
\newtheorem{example}[theorem]{Example}
\newtheorem{lemma}[theorem]{Lemma}
\newtheorem{corollary}[theorem]{Corollary}
\newtheorem{definition}[theorem]{Definition}

\newtheorem{conjecture}[theorem]{Conjecture}

\newcounter{countproblem}

\newcommand\numberthis{\addtocounter{equation}{1}\tag{\theequation}}

\newenvironment{problem}[1]
{
  \refstepcounter{countproblem}
  \begin{center}
    \begin{minipage}{0.98\linewidth}
      \begin{framed}
        {\bf Problem \arabic{countproblem} - {#1}.}
}
{  
      \end{framed}
    \end{minipage}
  \end{center}
}

\newenvironment{myitemize}{
\begin{itemize}
     \setlength{\itemsep}{0pt}
     \setlength{\parskip}{0pt}
     \setlength{\parsep}{-10pt}}
{\end{itemize}
}

\newenvironment{myenumerate}{
\begin{enumerate}
     \setlength{\itemsep}{0pt}
     \setlength{\parskip}{0pt}
     \setlength{\parsep}{-10pt}}
{\end{enumerate}
}

\newcommand{\pac}[1]{\left[#1\right]}

\newcommand{\Prob}{\mathbf P}

\newcommand{\polM}{\mathcal L_{\mathbf M}}
\newcommand{\polMh}{\mathcal L_{\mathbf M}^h}

\newcommand{\Z}{\mathbb{Z}}

\newcommand{\N}{\mathbb{N}}
\newcommand{\Q}{\mathbb{Q}}
\newcommand{\K}{\mathbb{K}}
\newcommand{\Kclos}{\overline{\mathbb{K}}}
\newcommand{\M}{\mathbf{M}}
\newcommand{\U}{\mathbf{U}}
\newcommand{\MS}{\mathbf{MS}}
\newcommand{\MNS}{\mathbf{MNS}}

\newcommand{\EV}{\mathbb{E}}
\DeclareMathOperator{\Var}{Var}
\newcommand{\Ni}{\mathcal{E}}

\newcommand{\mysection}[1]{\section{#1}}

\tikzset{%
  >=latex, 
}

\title{Computing Small Certificates of Inconsistency of \\Quadratic Fewnomial Systems}

\author[1]{Jean-Charles Faug\`ere}
\author[2]{Pierre-jean Spaenlehauer}
\author[3]{Jules Svartz}

\affil[1]{Sorbonne Universit\'es, UPMC, Univ. Paris 06, CNRS , INRIA, Laboratoire
d'Informatique de Paris 6 (LIP6), \'Equipe PolSys, 4 place Jussieu, 75252 Paris
Cedex 05, France. {\sf jean-charles.faugere@inria.fr}}

\affil[2]{Inria, CNRS, Universit\'e de Lorraine, Nancy, France.
  {\sf pierre-jean.spaenlehauer@inria.fr}}

\affil[3]{Minist\`ere de l'\'Education Nationale, Lyc\'ee Mass\'ena,
Sorbonne Universit\'es, UPMC,  
Univ. Paris 06, CNRS , INRIA, Laboratoire
d'Informatique de Paris 6 (LIP6), \'Equipe PolSys,
4 place Jussieu, 75252 Paris
Cedex 05, France.
{\sf jsvartz@ens-cachan.fr}
     }
\date{}

\begin{document}

\maketitle
\begin{abstract}
  B\'ezout's theorem states that dense generic systems of $n$
  multivariate quadratic equations in $n$ variables have $2^n$
  solutions over algebraically closed fields. When only a small subset
  $\M$ of monomials appear in the equations (\emph{fewnomial
    systems}), the number of solutions may decrease dramatically. We
  focus in this work on subsets of quadratic monomials $\M$ such that 
  generic systems with support $\M$ do not admit any solution at
  all. For these systems, Hilbert's Nullstellensatz ensures the
  existence of algebraic certificates of inconsistency. However, up to our knowledge all known bounds on
  the sizes of such certificates ---including those which take into
  account the Newton polytopes of the polynomials--- are exponential in
  $n$. Our main results show that if the inequality $2\lvert
  \M\rvert-2n \leq \sqrt{1+8\nu}-1$ holds for a quadratic fewnomial
  system -- where $\nu$ is the matching number of a graph associated
  with $\M$, and $\lvert\M\rvert$ is the cardinality of $\M$ -- then there exists generically a certificate of
  inconsistency of linear size (measured as the number of coefficients
  in the ground field $\K$). Moreover this certificate can be computed
  within a polynomial number of arithmetic operations.  Next, we evaluate how often this
  inequality holds, and we give evidence that the probability that the
  inequality is satisfied depends strongly on the number of
  squares. More precisely,
  we show that if $\M$ is picked uniformly at random among the
  subsets of $n+k+1$ quadratic monomials containing at least
  $\Omega(n^{1/2+\varepsilon})$ squares, then the probability that the
  inequality holds tends to $1$ as $n$ grows.  Interestingly, this phenomenon is related with the
  matching number of random graphs in the Erd\"os-Renyi model.
  Finally, we provide experimental results showing
  that certificates in inconsistency can be computed for systems with more than 
  10000 variables and equations.
\end{abstract}

\mysection{Introduction}

{\bf Context and problem statement.} Identifying classes of structured
polynomial systems and designing dedicated algorithms to solve them is
a central theme in computer algebra and in computational algebraic
geometry, due to the wide range of applications where such systems
appear.  We investigate quadratic systems involving a
small number of monomials (\emph{quadratic fewnomial systems}). Let
$\K$ be a field, $\Kclos$ its algebraic closure and $\M$ be a finite
subset of monomials of degree at most two in a polynomial ring
$\K[X_1,\ldots,X_n]$. Suppose also that the constant $1$ belongs to
$\M$, and let $\polM$ be the $\K$-linear space of polynomials in
$\K[X_1,\ldots, X_n]$ spanned by $\M$.

\smallskip

A classical question is to bound the number of solutions in $\Kclos^n$
of a system $f_1(X_1,\ldots,X_n)=\dots=f_n(X_1,\ldots, X_n)=0$, where
all polynomials lie in $\polM$ and have generic coefficients. When the
exponent vectors of the monomials in $\M$ are the points with integer
coordinates in a lattice polytope, Kushnirenko's theorem states that
the number of toric solutions (\emph{i.e.} solutions whose all
coordinates are nonzero) is bounded by the normalized volume of the
polytope~\cite{kushnirenko1976newton}. A variant of this theorem
indicates that such generic polynomial systems do not admit any
solution if the dimension of the $\Q$-linear space generated by the
exponent vectors of the monomials in $\M$ does not equal $n$.

\smallskip

Solving a polynomial system and deciding if it has any solution in
$\overline{\K}^n$ are two closely related questions. One classical
method to produce a certificate that a polynomial system does not have
any solution is to provide an algebraic relation via Hilbert's
Nullstellensatz:

\begin{problem}{Effective fewnomial Nullstellensatz}\label{pb:nullstellensatz}
  Given a system $(f_1,\ldots, f_m)\in \polM^m$ such that
  $f_1(X)=\dots=f_m(X)=0$ has no solution in $\overline{\mathbb K}^n$,
  compute $h_1,\ldots h_m\in \mathbb K[X_1,\ldots, X_n]$ such that
  $\sum_{i=1}^m f_i\,h_i = 1.$
\end{problem}

Bounding the sizes of the polynomials $h_i$ is a crucial question to estimate the complexity of this problem. In this paper, the notion of size that we use is the number of coefficients in $\K$ required to describe the polynomials $h_1,\ldots, h_m$.

On the other hand, the specification of ``solving a polynomial
system'' depends on the context. If the number of solutions in the
algebraic closure $\overline{\K}$ is finite, one way to represent them
symbolically is to provide a rational parametrization of their
coordinates by the roots of a univariate polynomial. For the sake of simplicity, we consider only the problem of computing
a univariate polynomial whose roots parametrize the set of solutions:

\begin{problem}{Partial $0$-dimensional fewnomial system solving}\label{pb:solving}
Given a polynomial system $f_1=\dots=f_m=0$ with support $\M$ that
have finitely-many solutions in $\overline{\mathbb K}^n$ and a
monomial $\mu\in\M$, compute a univariate polynomial $P_\mu\in \mathbb
K[\mu]$ which vanishes at all the solutions of the system.
\end{problem}

Hence, the roots of the univariate polynomial $P_\mu$ contain the
images of the solutions of the input multivariate system via the
monomial map $(X_1,\ldots, X_n)\mapsto \mu$. 

\medskip

{\bf Related works.}  Sparse elimination theory for solving systems
with special monomial structures have been developed since the 80s
\cite{Stu91}. Several lines of work have been initiated during this
period. When the exponent vectors of the monomials occurring in the
polynomials of the system are the lattice points in a lattice
polytope, connections with convex and toric geometry have been
established and dedicated solving methods have been designed: homotopy
continuation methods
\cite{huber1995polyhedral,verschelde1994homotopies}, resultants
\cite{Stu91, canny2000subdivision}, Gr\"obner bases
\cite{Stu96,FauSpaSva14}, etc. One important theme of these
developments is to relate algebraic structures with combinatorial
properties of convex bodies. In particular, Kushnirenko and
Bernshtein's theorems
\cite{kushnirenko1976newton,bernshtein1975number} provide bounds on
the number of isolated toric solutions in terms of the Newton
polytopes of the input polynomials.  Another line of work have been
initiated by Khovanskii in the 80s on fewnomial systems
\cite{khovanskii1980class}. The main theme in this setting is to
relate the algebraic and algorithmic complexity of several problems to
the number of monomials occurring in the equations. For instance, a
classical and challenging question is to bound the number of real
solutions in the positive orthant, see \emph{e.g.}
\cite{khovanskii1980class,shub1996intractability,bertrand2006polynomial,bihan2007new,DBLP:journals/corr/KoiranPT13,DBLP:journals/corr/KoiranPTT13}
for results on this topic.  Bounding the size of a certificate of
inconsistency of a polynomial system via the Nullstellensatz is a
classical problem.  Up to our knowledge, all known upper bounds on the size
of such certificates are exponential in the number of variables $n$;
moreover, examples by Masser and Philippon and by Lazard and Mora show
that one cannot hope for better bounds in the worst case. A classical
bound is given by Kollar \cite{kollar1988sharp}: if the maximal degree
of the input inconsistent system $f_1,\ldots, f_m$ is $D$, then there
exist $h_1,\ldots, h_m$ such that $\sum_{i=1}^m f_i\, h_i=1$ and the
degrees of the $h_i$ are bounded by
$n\min(n,m)D^{\min(n,m)}+\min(n,m)D$. This bound is general and does
not require any further assumption. It was later improved to $\deg(f_i
h_i)\leq \max(3,D)^n$ \cite{Fit}. When there is no solution at
infinity, the degrees of the polynomials $h_i$ are bounded by $(D-1)n$
\cite{Laz77,brownawell1987bounds}: the number of coefficients in dense
polynomials of this degree is still exponential in $n$. For general polynomial
systems, it would be surprising that certificates of inconsistency with size polynomial
in the input size exist, as this would imply $NP=coNP$. Estimates
taking into account the bitsize of the coefficients that appear in the
certificate in terms of the bitsize of the coefficients of the input
system are provided by \emph{Arithmetic Nullstellens\"atze}, see
\emph{e.g.}  \cite{krick2001sharp} and references within.  Two
milestones on the sparse effective Nullstellensatz are the bounds in
\cite{canny2000subdivision} and \cite{sombra1999sparse}: these bounds
provide certificates of size bounded by a quantity which depends on
the Newton polytopes of the input polynomials. However, their size is
exponential in the size of the input, and these bounds do not take
into account the sparsity of the support inside its Newton
polytope. One of the main difficulty to generalize these techniques to
fewnomial quadratic systems is the fact that the proofs rely on algebraic
properties (normality, Cohen-Macaulay algebras) that hold for
semigroup algebras generated by lattice points in normal polytopes,
but not for semigroup algebras generated by a scattered set of
monomials.

Other models of sparse systems have also been investigated. For
instance, systems where each quadratic equation involves a small
subset of variables have been investigated in \cite{semaev2008solving}
and \cite{DBLP:journals/corr/CifuentesP14}.
Connections between combinatorial properties of graphs and polynomial
systems is a classical topic which has been investigated from several
viewpoints. For instance, square-free monomial ideals have many
combinatorial properties and can be seen as the \emph{edge ideals} of
graphs, see \emph{e.g.}
\cite{froberg1990stanley,herzog2003monomial}.
Connections between the regularity of the edge ideal of a graph and
its matching number and co-chordal cover number are shown in
\cite{Woo10}. Cohen-Macaulay criteria for such ideals are investigated
in \cite{herzog2005distributive,crupi2011cohen}.

Bounds on the size of certificate of inconsistency of polynomial systems are a
important ingredient in algebraic proof complexity, see \emph{e.g.} the
Nullstellensatz proof system \cite{beame1994lower} and related works \cite{clegg1996using}.

\medskip

{\bf Main results.} An open question is whether there exist
certificates of inconsistency of polynomial size for general fewnomial
systems involving $n+k+1$ monomials in $n$ variables. The goal of this
work is to investigate this question in the case of quadratic
polynomials. We present an explicit criterion which identifies subsets
$\M$ of monomials of degree at most $2$ such that systems of $n$
equations in $n$ variables with support $\M$ and generic coefficients
do not have any solution, and such that there exists a sparse
certificate $\sum_{i=1}^n f_i h_i=1$, where all polynomials
$h_1,\ldots, h_n$ lie in $\polM$.  Therefore, the number of
coefficients in $\K$ required to represent the certificate is the same
as that of the input system.  Moreover, when $\M$ is such a subset, we
propose a method which computes such $h_1,\ldots, h_n$ within
a polynomial number of arithmetic operations.

More precisely, we model the set $\M$ by a graph $G$ on $n+1$
vertices, where each edge represents a nonconstant square-free monomial
in $\M$. The constant $1$ and the squares in $\M$ are
distinguished with loops in the graph (the precise construction is
described in Section~\ref{sec:graph}). Let $\nu(\M)$ denote the
\emph{matching number} (\emph{i.e.} the maximum cardinality of a matching) of the subgraph of vertices in $G$ with a loop. 

\begin{theorem}\label{thm:match}
  If $m\geq \lvert\M\rvert-\frac{\sqrt{1+8\nu(\M)}-1}{2}$, then a generic
  system $(f_1,\ldots, f_m)\in\polM^m$ has no solution in
  $\overline{\mathbb K}^n$. Moreover, there exists $(h_1,\ldots, h_m)$
  solving Problem \ref{pb:nullstellensatz} s.t. all $h_i$ lie in
  $\polM$ and they can be computed within $O\left(m^\omega
  \binom{\lvert\M\rvert+1}{2}^\omega\right)$ operations in $\K$, where $\omega$
  is a feasible exponent for matrix multiplication ($\omega<2.37286$
  with Le~Gall's algorithm \cite{Leg14}).
\end{theorem}

We would like to emphasize that the inequality $m\geq
\lvert\M\rvert-\frac{\sqrt{1+8\nu(\M)}-1}{2}$ can be checked in
polynomial time, since the matching number of a graph can be computed
in polynomial time with Edmonds' algorithm~\cite{edmonds1965paths}.
Next, we relate how often the assumptions of Theorem \ref{thm:match}
hold with the number of squares in the support $\M$.
If the subset of
square monomials and the subset of square-free monomials in $\M$ are
chosen at random, and the cardinality of $\M$ is $n+k+1$ and the number of
squares is larger than $\Omega(n^{1/2+\varepsilon})$ for some
$\varepsilon>0$, then the assumptions of Theorem \ref{thm:match} hold with
large probability, leading to the following statement:

\begin{theorem}\label{thm:introrandgraph}
  Let $k$ be a fixed integer, $a_n,b_n\in\N$ be such $a_n+b_n=n+k+1$,
  and $\M$ be a subset of monomials of degree at most $2$
  in $\K[X_1,\ldots, X_n]$ distributed uniformly at random among those
  that contain the constant $1$, $a_n$ nonsquare monomials, and
  $b_n-1$ non-constant square monomials. Assume further that
  $b_n=\Omega(n^{1/2+\varepsilon})$, for $\varepsilon>0$.  Then the
  probability that the assumptions of Theorem
  \ref{thm:match} with $m=n$ are satisfied for
  $\M$ tends towards $1$ as $n$ grows.
\end{theorem}

The cornerstones of the proof of this theorem rely on properties of
random graphs in the Erd\"os-Renyi model.  Experiments suggest that
this result is sharp: when there are at most $O(n^{1/2})$ square
monomials in $\M$, we observe experimentally that the probability of
having a certificate in $\polM^m$ seems to converge to a non-zero value smaller than $1$ as $n$ grows. This is also the case when the support is chosen uniformly at random (the expected number of squares is $O(1)$). We propose a conjecture stating that the limit probability is nonzero in that case.

We also study a limit case: when $\lvert\M\rvert=n+k+1$ and all the
squares are in $\M$. The generic number of solutions in this setting
is given by the B\'ezout's theorem: it equals $2^n$. We
shall see that with probability tending to~1, these solutions can be
compactly represented as the orbits of $2^{2k+2}$ points under an
action of $(\Z/2\Z)^{n-2k-2}$. Computing these solutions amounts to
solving a system of $2k+2$ equations in $2k+2$ variables: the
time complexity of this task does not depend on $n$. A direct consequence
is that computing a compact representation of the solutions of such
systems require a number of operations in $\K$ which is polynomial in
$n$, even though their number of solutions is exponential in $n$.
This suggests that the number of solutions in the algebraic closure, which is often used to measure the complexity of solving polynomial systems,
might in some cases greatly overestimate the complexity
for fewnomial systems.  Another open issue is to extend this work to the
non quadratic case.

\smallskip

Finally, we show experimental results obtained with our proof-of-concept
implementation. They show that certificates of inconsistency can be computed
for quadratic fewnomial systems with more than 10000 variables
and equations when there are sufficiently many squares in the monomial support.
Moreover, we also observe some unexpected behaviors which raise new questions about
fewnomial systems. For instance, as $n$ grows, there seems to be a phase
transition in the probability of having a small certificate of inconsistency.
Moreover, in the case where there are few squares in the fewnomial system (this
case is not covered by the theoretical analysis), there seems to be a non-zero
probability that a fewnomial system has a small certificate of inconsistency.
These phenomenons remain to be explained.

\medskip

{\bf Organization of the paper.} Section \ref{sec:notation} introduces
notation and states preliminary results.
The core result of the paper is proved in Section \ref{sec:graph},
establishing a connection between the matching number and the
existence of a small certificate of inconsistency. Section
\ref{sec:randomgraph} is devoted to a probabilistic analysis of the
matching number of some random graphs in the Erd\"os-Renyi
model. Section \ref{sec:allsquares} investigates some families of
fewnomial systems where all squares appear in the equations. Finally,
we report experimental results in Section \ref{sec:experiments} and
state a conjecture for quadratic fewnomial systems involving few square
monomials.

\mysection{Notation and preliminaries} \label{sec:notation}

{\bf Notation.} Throughout this paper, $\K$ denotes a field of odd characteristic. Its
algebraic closure is denoted by $\overline{\K}$. If $X_1,\ldots, X_n$ are
variables, and $\alpha\in \N^n$, then the shorthand $X^\alpha$ stands
for the monomial $X_1^{\alpha_1}\dots X_n^{\alpha_n}$. The symbol $\M$
denotes a finite subset of monomials in $\K[X_1,\ldots, X_n]$
containing the constant $1$. For any $i\in\N$, $\M^i$ denotes the
subset of all products of $i$ monomials in $\M$. Its cardinality is
denoted by $\lvert\M^i\rvert$. By convention, $\lvert\M^0\rvert=1.$ By
slight abuse of notation, we call \emph{dimension} of an ideal $I$ in
a ring $R$ the Krull dimension of the quotient ring $R/I$.

\smallskip

{\bf Complexity model.} Complexity bounds in this paper count the
number of operations $\{+,-,\times,\div\}$ in the field $\K$. It is
not our goal to take into account the bitsize of the
coefficients in $\K$. Hence, we count each arithmetic operation with
unit cost. We do not take into account operations on monomials. The
notion of size that we use for polynomial systems is the number of
coefficients in $\K$ required to represent them. Note that if $\K$ is
a finite field, then the bitsizes of the elements in $\K$ are bounded, and
hence the bit complexity is the same as the arithmetic complexity.
Given partial functions $g, h$ from a set $I$ to $\N$, we use the
following classical Landau notation: $f=O(g)$ means that $f/g$ is
bounded above by a constant, $f=\Omega(g)$ is equivalent to $g=O(f)$,
and $f=\Theta(g)$ means that $f=O(g)$ and $g=O(f)$.

\smallskip

{\bf Genericity.} Let $\polM$ denote the $\K$-linear space spanned by $\M$. It
has dimension $\lvert \M\rvert$. We say that a property holds for a generic
system $(f_1,\ldots, f_m)\in \polM^m$ if there exists a dense Zariski open
subset $\mathcal O$ of $\polM^m$ s.t. this property holds for any system in
$\mathcal O$.

\smallskip

{\bf Semigroup algebras.} The main algebraic structure that we
consider are \emph{semigroup algebras} (also called \emph{toric
  rings}): if $\M\subset \K[X_1,\ldots, X_n]$ is a finite subset of
monomials containing $1$, we let $\K[\M]$ denote the subalgebra of
$\K[X_1,\ldots, X_n]$ generated by $\M$. We do not make any assumption
on the Krull dimension of the ring $\K[\M]$. Semigroup algebras which are domains are
the coordinate rings of affine toric varieties \cite{CoxLitSch11}. We
refer to \cite[Ch. 7]{MilStu05} for a more detailed presentation. By slight abuse of notation, we call
variety of a system $f_1,\ldots, f_m\in\K[\M]$ the variety in
$\overline{\K}^n$ associated to the ideal $\langle f_1,\ldots,
f_n\rangle\subset\K[X_1,\ldots, X_n]$.

\smallskip

The following proposition is a variant of the weak Nullstellensatz for the total coordinate
ring of projective toric varieties (see \emph{e.g.} \cite[Prop.
5.2.6]{CoxLitSch11}). 
\begin{proposition}\label{prop:toricNullstellesatz}
The variety associated with a system $f_1,\ldots, f_m\in\K[\M]$ is empty if and only if
there exist $h_1,\ldots, h_m\in\K[\M]$ such that
$\sum_{i=1}^m f_i\, h_i = 1.$
\end{proposition}

\begin{proof}
  the ring $\K[\M]$ is isomorphic to $\K[X]/I_M$, where $I_M$ is a toric ideal
  generated by binomials $b_1,\ldots, b_\ell$. Let $\widetilde{f_1},\ldots,
  \widetilde {f_m}$ be the images of $f_1,\ldots, f_m$ by the isomorphism. Using the Nullstellensatz on the
  system $\widetilde{f_1},\ldots, \widetilde{f_m}, b_1,\ldots, b_\ell$ in
  $\K[X]$ and pulling it back to $\K[\M]$ proves the
  proposition.
\end{proof}

Proposition \ref{prop:toricNullstellesatz} indicates that we can look for polynomial relations in
$\K[\M]$ instead of the whole algebra $\K[X]$. Although narrowing the search
for the certificate in $\K[\M]$ instead of $\K[X]$ constrains the problem, we shall see that this approach enables us to find efficiently small certificates. This leads to the following variant of Problem \ref{pb:nullstellensatz}:

\begin{problem}{Effective fewnomial Nullstellensatz in $\K[\M]$}\label{pb:nullstellkM}
Given a system $f_1,\ldots, f_m\in \mathbb K[\M]$
and such that $f_1(X)=\dots=f_m(X)=0$
has no solution in $\overline{\mathbb K}^n$, compute $h_1,\ldots,h_m\in \mathbb K[\M]$ such that
$\sum_{i=1}^m f_i\,h_i = 1.$
\end{problem}

\mysection{Monomials and support graphs}
\label{sec:graph}

In this section, we show a connection between graphs and properties of
$\K[\M]$.  In particular, we focus on quadratic relations between
monomials in $\M$, \emph{i.e.} at $\K$-linear relations in the vector
space spanned by $\M^2$.  We start by adding a new variable $X_0$ and
by considering the homogenized support
$\M^h=\{X_0^{2-\deg(\mu)}\mu\}_{\mu\in\M}$.  We associate with $\M$ a
simple labeled undirected graph $G$ on $S=\{0,\ldots, n\}$ whose edges
are $E=\{ (i,j) \mid X_i X_j\in \M^h, i\ne j\}$.  There is a loop at
a vertex $i$ iff $X_i^2\in\M^h$.

\begin{example}
Let $\M=\{1,X_1^2,X_2^2, X_3^2, X_3, X_4, X_1 X_2,  X_2 X_3, X_3X_4\}$. The following picture represents the graph $G$; squares in $\M^h$ are indicated by a loop.

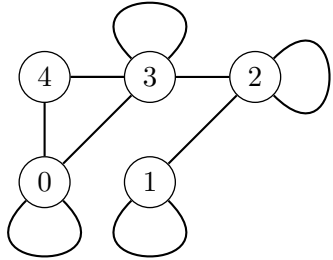
\begin{center}
\begin{tikzpicture}[scale=0.7]
\GraphInit[vstyle=Normal]
\SetGraphUnit{2}
\Vertex[L=$0$]{1}
\EA[L=$1$](1){2}
\NOEA[L=$2$](2){3}
\WE[L=$3$](3){4}
\WE[L=$4$](4){5}
\Loop[dist=2cm,dir=SO,style={thick}](1)
\Loop[dist=2cm,dir=SO,style={thick}](2)
\Loop[dist=2cm,dir=EA,style={thick}](3)
\Loop[dist=2cm,dir=NO,style={thick}](4)
\Edges(2,3,4,5,4,1,5)
\end{tikzpicture}

\end{center}
\end{example}

\begin{figure}\centering
\begin{minipage}[t]{0.3\linewidth}
\begin{center}Type 1:\\{\small$(X_i X_j)\cdot(X_k X_\ell)=(X_i X_\ell)\cdot(X_j X_k)$}

  \begin{tikzpicture}[scale=0.5]
    \GraphInit[vstyle=Normal]
    \SetGraphUnit{2}
    \Vertex[L=$i$]{1}
    \NOEA[L=$j$](1){2}
    \SOEA[L=$k$](2){3}
    \SOEA[L=$\ell$](1){4}
    \Edges(1,2,3,4,1)
\end{tikzpicture}\end{center}
\end{minipage}
\quad
\begin{minipage}[t]{0.3\linewidth}
\begin{center}Type 2:\\{\small$(X_i X_j)\cdot(X_i X_k)=(X_i^2)\cdot(X_j X_k)$}

\begin{tikzpicture}[scale=0.5]
\GraphInit[vstyle=Normal]
\SetGraphUnit{2}
\Vertex[L=$i$]{1}
\Loop[dist=2cm,dir=WE,style={thick}](1)
\NOEA[L=$j$](1){2}
\SOEA[L=$k$](1){3}
\Edges(2,1,3,2)
\end{tikzpicture}
\end{center}
\end{minipage}
\quad
\begin{minipage}[t]{0.3\linewidth}
\begin{center}Type 3:\\{\small$(X_i X_j)\cdot(X_i X_j)=(X_i^2)\cdot(X_j^2)$}

\begin{tikzpicture}[scale=0.5]
\GraphInit[vstyle=Normal]
\SetGraphUnit{3}
\Vertex[L=$i$]{1}
\EA[L=$j$](1){2}
\Loop[dist=2cm,dir=NO,style={thick}](1)
\Loop[dist=2cm,dir=SO,style={thick}](2)
\Edges(1,2)
\end{tikzpicture}
\end{center}
\end{minipage}
\caption{The three types of quadratic relations \label{fig:toricrel}}
\end{figure}
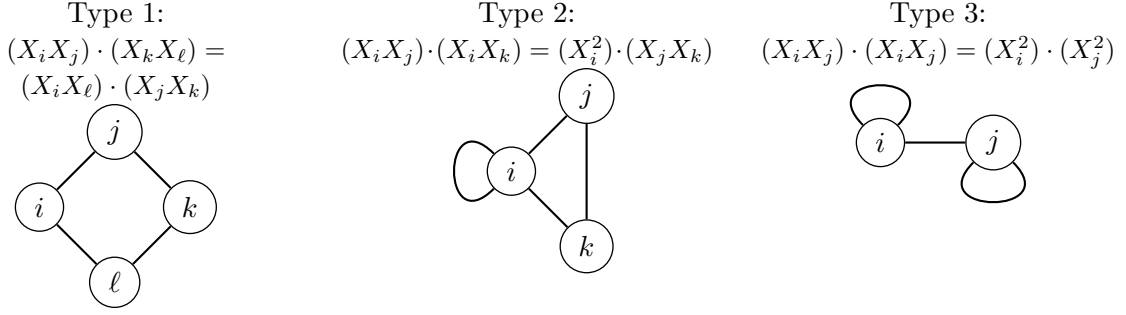

Quadratic relations between elements of
$\M$ are of the form $\mu_1\, \mu_2 = \mu_3\, \mu_4$ for (not
necessarily distinct) monomials $\mu_1,\mu_2,\mu_3,\mu_4\in \M$.  For the quadratic supports $\M$ that we
consider in this paper, these quadratic relations come in three flavors that appear as subgraphs of 
$G$ and are described in Figure~\ref{fig:toricrel}. The next proposition shows
how the cardinality of $\M^2$ can be computed from the number of quadratic
relations and the number of $4$-cliques in $G$. We recall that a $4$-clique is
a subgraph on $4$ vertices such that every pair of vertices is linked by an
edge.

\begin{proposition}\label{prop:cardM2}
The cardinality of $\M^2$ equals
$\binom{\lvert\M\rvert+1}{2}-\lambda(G)+{\rm clique}_4(G)$,
where $\lambda(G)$ is the number of subgraphs of $G$ isomorphic to any of the three graphs in Figure \ref{fig:toricrel} and ${\rm clique}_4(G)$ is the number of 4-cliques in $G$.
\end{proposition}
\noindent\emph{Proof.} We can form $\binom{\lvert\M\rvert+1}{2}$ products of two (non-necessarily
distinct) elements in $\M$. However, some of these products are counted
several times because of the quadratic relations between elements in $\M$. This is corrected by the terms $-\lambda(G)+{\rm clique}_4(G)$; we detail below the possible cases: 
\begin{myitemize}
\item If $\mu=X_iX_jX_kX_\ell$ is a product of four distinct variables, then
  $\mu$ can be obtained from $\M$ by three different products, since
  $\mu=(X_iX_j)(X_kX_\ell)=(X_iX_k)(X_jX_\ell)=(X_iX_\ell)(X_jX_k)$. Depending
  on the number of pairs of such edges that lie in the graph, the monomial
  $\mu$ is counted one, two or three times in $\binom{\lvert\M\rvert+1}{2}$.\\
  \noindent If there is only one way to obtain $\mu$ (for example if $(X_iX_j)$
  and $(X_kX_\ell)$ are the only monomials in $\M$ whose products are $\mu$),
  then the subgraph associated with the vertices $\{X_i,X_j,X_k,X_\ell\}$ is
  neither of type 1 nor a 4-clique. Hence, $\mu$ is counted only one time.\\
  \noindent If there are two ways to obtain $\mu$, then the subgraph associated with
  the vertices $\{X_i,X_j,X_k,X_\ell\}$ is of type 1 but not a 4-clique. Hence,
  $\mu$ is counted twice in $\binom{\lvert\M\rvert+1}{2}$ but this is corrected
  by the term $\lambda(G)$.\\
  \noindent If all the three products are possible, then the subgraph associated with the vertices $\{X_i,X_j,X_k,X_\ell\}$ contains three subgraphs of type 1, and is also a 4-clique. Therefore $\mu$ is counted $3$ times in $\binom{\lvert\M\rvert+1}{2}$, removed $3$ times in $\lambda(G)$ and counted once in ${\rm clique}_4(G)$.
\item If $\mu=X_i^2X_jX_k$ is a monomial involving three distinct variables, then $\mu$ is counted twice in $\binom{\lvert\M\rvert+1}{2}$ if and only the subgraph associated with the vertices $\{X_i,X_j,X_k\}$ is of type 2. In this case one contribution is removed by the term $\lambda(G)$, hence $\mu$ is counted one time.
\item Similarly, monomials $\mu=X_i^2X_j^2$ are counted once or twice in the
  formula $\binom{\lvert\M\rvert+1}{2}$: if it is counted twice (\emph{i.e.}
  when $X_iX_j, X_i^2, X_j^2\in\M$), then the subgraph associated with $\{X_i,X_j\}$ is of type 3. \hfill$\square$
\end{myitemize}
{\bf Notation.} For a graph $G$ associated with a set of monomials $\M$, let $G'$ be the subgraph of squares (\emph{i.e.} the subgraph of vertices with a loop). 
\begin{definition}\label{def:matching}
A \emph{matching} (also called independent edge set) of $G'$ is a set of edges of $G'$ without common vertices. We let $\nu(\M)$ denote the \emph{matching number} of $G'$, \emph{i.e.} the maximum cardinality of a matching of $G'$.
\end{definition}

The matching number of a graph can be computed in polynomial time by Edmonds's algorithm \cite{edmonds1965paths}.
We refer to \cite{lovasz1986matching} for more details on matching theory.
We state now the main result of this section, which connects the matching number of the graph $G'$ to the existence of a small certificate of inconsistency:
\begin{theorem}\label{thm:semiregfriendly}
Let $(f_1,\ldots, f_m)\in\polM^m$ be a system with gene\-ric coefficients.
If $m\geq \lvert\M\rvert-\frac{\sqrt{1+8\nu(\M)}-1}{2}$, then there exist
polynomials $h_1,\ldots, h_m\in\polM$ such that 
$\sum_{i=1}^m f_i h_i =1.$
\end{theorem}
The proof of this theorem is postponed to the end of the section. It is
actually not surprising that systems satisfying the assumptions of Theorem \ref{thm:semiregfriendly} do not have any solution, since the dimension of the $\Q$-vector space generated by the exponent vectors in $\M$ is upper bounded by $\lvert \M\rvert-\nu(\M)$: each edge $(i,j)$ in $G'$ means that $X_i^2, X_j^2, X_i X_j\in\M$ and the exponent vectors of these three monomials are linearly dependent over $\Q$.
The main point of Theorem \ref{thm:semiregfriendly} is that, under the condition on $\nu(\M)$, the polynomials $(h_1,\ldots, h_m)$ for the effective Nullstellensatz can be searched in $\polM$. This allows to obtain to get small certificates of inconsistency:
\begin{corollary}\label{cor:complNullstellensatz}
  With the notation and under the assumptions of Theorem
  \ref{thm:semiregfriendly}, there is an explicit algorithm which solves
  Problem \ref{pb:nullstellensatz} within
  $O\left(m\lvert\M\rvert\left(\binom{\lvert\M\rvert+1}{2}-\lambda(G)+{\rm
        clique}_4(G)\right)^{\omega-1}\right)$ arithmetic operations, where $\omega$ is a
  feasible exponent for matrix multiplication ($\omega<2.37286$ with
  Le~Gall's algorithm \cite{Leg14}). This complexity is polynomial in the number of coefficients $m\lvert\M\rvert$ of the input system.
\end{corollary}

\begin{proof}
  According to Theorem \ref{thm:semiregfriendly}, there exist
  polynomials $h_1,\ldots, h_m$ with support $\M$ s.t. $\sum_i h_i f_i
  =1$, by decomposing the polynomials $h_i$ in the monomial basis
  there exists a relation $$\sum_i \sum_{\mu\in \M}
  \alpha_{\mu,i}\, \mu f_i = 1$$ where $\alpha_{\mu,i}\in\K$.  Let
  $V\subset{\sf Span}_\K(\M^2)$ be the linear space generated by the
  products $\{\mu\,f_i\}_{\mu\in\M,i\in\{1,\ldots,
    m\}}$. Consequently, computing the polynomials $h_i$ amounts to
  solving a linear system over $\K$ with $m\,\lvert\M\rvert$ unknowns
  and $\lvert\M^2\rvert$ equations. Solving it requires
  $O(m\lvert\M\rvert\cdot \lvert\M^2\rvert^{\omega-1})$ operations in
  $\K$ \cite[Prop.~2.11]{storjohann2000algorithms}. Proposition
  \ref{prop:cardM2} concludes the proof. 
\end{proof}

The sequel of this section is devoted to this proof of
Theorem~\ref{thm:semiregfriendly}. The squareroot involved in the formula is a
consequence of the following lemma, as the maximal value of $p$ for which
$n\geq \binom{n-p+1}2$.
\begin{lemma}\label{lem:quadsquare}
There exist linear forms $\ell_1,\ldots, \ell_p\in \K[X_1,\ldots, X_n]$ such that the ideal
$$I=\langle X_1^2,\ldots, X_n^2, \ell_1(X_1,\ldots, X_n),\ldots,
\ell_p(X_1,\ldots, X_n)\rangle$$ contains all monomials of degree $2$
iff $p\geq n- \frac{\sqrt{1+8 n}-1}{2}$.
\end{lemma}
\begin{proof}
The vector space of $(n-p)$-variate quadratic forms has dimension $\binom{n-p+1}{2}$. From the inequality $p\geq n- \frac{\sqrt{1+8 n}-1}{2}$, we obtain $n\geq \binom{n-p+1}{2}$. This inequality and the fact that any quadratic form can be written as a linear combination of squares of linear forms (since ${\rm char}(\K)\ne 2$), implies that there exist $\ell'_1,\ldots, \ell'_n$ such that their squares ${\ell'_1}^2,\ldots, {\ell'_n}^2$ generate the space of $(n-p)$-variate quadratic forms.
Then the dimension of the linear space generated by $\ell'_1,\ldots, \ell'_n$ is necessarily maximal and equals $n-p$. Up to permuting the indices, we assume also that $\ell'_1,\ldots, \ell'_{n-p}$ are linearly independent.
Hence the ideal
$I'=\langle {\ell'_1}(X_1,\ldots,X_{n-p})^2,\ldots,
{\ell'_n}(X_1,\ldots,X_{n-p})^2, X_{n-p+1},\ldots, X_n\rangle$
contains all monomials of degree $2$. We rewrite $I'$ as
$$I'=\langle {\ell''_1}(X_1,\ldots,X_n)^2,\ldots,
{\ell''_{n}}(X_1,\ldots,X_{n})^2, X_{n-p+1},\ldots,
X_n\rangle,
$$ 
$$\quad\quad\begin{cases}
\ell''_i(X_1,\ldots, X_n)=\ell'_i(X_1,\ldots, X_{n-p})\text{ if } 1\leq i\leq n-p\\
\ell''_i(X_1,\ldots, X_n)=X_i-\ell'_i(X_1,\ldots, X_{n-p})\text{ otherwise}.
\end{cases}$$

Note that the linear forms $\ell''_1,\ldots,\ell''_n$ are linearly independent by construction.
We consider the automorphism $\theta$ of $\K[X_1,\ldots, X_n]$ defined by $\theta(X_i)=\ell''_i(X_1,\ldots, X_{n})$,
and we set $\ell_i(X_1,\ldots, X_n)=\theta^{-1}(X_{n-p+i})$ for
$i\in\{1,\ldots, p\}$. Therefore $I$
is the inverse image of $I'$ by $\theta$ and hence contains all the monomials of degree $2$.

It remains to prove the converse statement, \emph{i.e.} that $p< n- \frac{\sqrt{1+8 n}-1}{2}$ implies that there do not exist such linear forms $\ell_1,\ldots, \ell_p$. This is achieved by a similar argument: if such linear forms existed, then there would exist a set of $n$ generators of the vector space of $(n-p)$-variate quadratic forms. This is not possible if $p< n- \frac{\sqrt{1+8 n}-1}{2}$ since this vector space has dimension $\binom{n-p+1}{2}$.
\end{proof}

We can now prove the main theorem of this section:
\begin{proof}[Proof of Theorem \ref{thm:semiregfriendly}]
We prove a homogeneous version of Theorem \ref{thm:semiregfriendly}: let $f_1^{(h)},\ldots, f_m^{(h)}\in\polMh\subset\K[X_0,\ldots, X_n]$ be the homogenization of the generic system $f_1,\ldots, f_m$. We shall show that any monomial in $(\M^h)^2$ (see the definition of $\M^h$ at the beginning of this section) belongs to the ideal $\langle f_1^{(h)},\ldots, f_m^{(h)} \rangle\subset \K[\M^h]$. This will imply that there exist $h_1^{(h)},\ldots, h_m^{(h)}\in \K[\M^h]$ such that
$\sum_{i=1}^m f_i^{(h)} h_i^{(h)}=X_0^4\in(\M^h)^2.$
Setting $X_0=1$ in this equation yields the desired relation.

First, we prove the existence of one system $f_1^{(h)},\ldots, f_m^{(h)}$ such that
all monomials of $(\M^h)^2$ appear in the ideal $\langle f_1^{(h)},\ldots,
f_m^{(h)}\rangle$.  Throughout this proof, we let 
$A=\{\{a_1,b_1\},\ldots,
\{a_{\nu(\M)},b_{\nu(\M)}\}\}\subset \{0,\ldots, n\}^2$ denote a matching of $G'$ of maximum
cardinality. We construct a system from $A$ whose polynomials are:
\begin{myenumerate}
\item all the monomials in $\M^{h}$ of the form $X_i X_j$ with $i\neq j$;
\item all the monomials in $\M^h$ of the form $X_i^2$ with $i$ not appearing in $A$;
\item for each $i\in \{1,\ldots,\nu(\M)\}$, the polynomial $X_{a_i}^2-X_{b_i}^2$;
\item the polynomials $\ell_1(X_{a_1}^2,\ldots,X_{a_{\nu(\M)}}^2),\ldots,
  \ell_p(X_{a_1}^2,\ldots,X_{a_{\nu(\M)}}^2)$, where the $\nu(\M)$-variate linear forms $\ell_1,\ldots,\ell_p$  
are obtained by replacing $n$ by $\nu(\M)$ in Lemma \ref{lem:quadsquare}.
\end{myenumerate}
This is a system of $\lvert\M\rvert-\left\lfloor\frac{\sqrt{1+8\nu(\M)}-1}{2}\right\rfloor$ polynomials, generating an ideal $I\subset \K[\M^h]$.
We claim that all monomials in $(\M^h)^2$ are in the ideal of $\K[\M^h]$ generated by these polynomials:
\begin{myitemize}
\item every monomial in $(\M^h)^2$ involving at least $3$ different variables belongs necessarily to the ideal generated by the monomials $X_i X_j$ with $i\neq j$; the same holds for monomials of the form $X_i^3 X_j$ with $i\neq j$;
\item next, we look at monomials of the form $X_i^4$. If $i$ does not appear in $A$, then by construction $X_i^2$ is in the ideal. If $i$ is in $A$, then there exists $j$ such that $i=a_j$ or $i=b_j$. Noticing that $X_{a_j}^4=X_{a_j}^2(X_{a_j}^2-X_{b_j}^2) - (X_{a_j} X_{b_j})^2$ or $X_{b_j}^4=X_{b_j}^2(X_{a_j}^2-X_{b_j}^2) - (X_{a_j} X_{b_j})^2$ shows that $X_i^4\in I$.
\item finally, we focus on monomials of the form $X_i^2 X_j^2$. If $i$ or $j$
  do not appear in $A$, then either $X_i^2$ or $X_j^2$ belongs to $I$. If both
  $i$ and $j$ appear in $A$ then Lemma \ref{lem:quadsquare} tells us that
  $X_i^2 X_j^2$ belongs to the ideal generated by $\langle X_{a_1}^4,\ldots,
  X_{a_{\nu(\M)}}^4,\ell_1(X_{a_1}^2,\ldots,X_{a_{\nu(\M)}}^2),\ldots,
  \ell_p(X_{a_1}^2,\ldots,X_{a_{\nu(\M)}}^2),X_{a_1}^2-X_{b_1}^2,\ldots, X_{a_{\nu(\M)}}^2-X_{b_{\nu(\M)}}^2\rangle$.
\end{myitemize}

So far, we have proven that there exists at least one system such that
Theorem \ref{thm:semiregfriendly} is correct. It remains to prove that
this is true for a generic system. To this end, we note that all
monomials in $(\M^h)^2$ belongs to $\langle f_1^{(h)},\ldots,
f_m^{(h)}\rangle\subset\K[\M^h]$ if and only if ${\sf Span}_\K (\{\mu\,
f^{(h)}_i\}_{\mu\in\M^h,i\in\{1,\ldots, m\}})={\sf Span}_\K((\M^h)^2)$. This is an
open condition given by the non-vanishing of a product of minors of
the matrix recording the coefficients of $\{\mu\,
f_i^{(h)}\}_{\mu\in\M,i\in\{1,\ldots, m\}}$. Consequently, there exists a
Zariski open subset $\mathcal O\subset \K[\M^h]^m$ such that Theorem
\ref{thm:semiregfriendly} holds. This open subset $\mathcal O$ is
non-empty by the construction above. The proof is concluded by
noticing that any non-empty open subset is dense in the Zariski
topology.
\end{proof}

\mysection{Random support graphs}\label{sec:randomgraph}
In this section, we assume that the support $\M$ is randomly generated
and we estimate the probability that the assumptions of Theorem
\ref{thm:semiregfriendly} are satisfied.  Roughly speaking, the aim of this section
is to show that the conditions of Theorem \ref{thm:semiregfriendly} hold with large
probability if $n$ is large enough and if there are sufficiently many
squares in $\M$.
Let us consider the following variant of the Erd\"os-R\'enyi random graph model: for
$n\in\N$, we set two probabilities $p_n,q_n\in[0,1]$, and we consider
a sequence of random supports $(\M_n)_{n\in\N}$ where
\begin{myitemize}
\item $\M_n$ is a subset of quadratic monomials of $\K[X_0,\ldots, X_n]$; 
\item each square $X_i^2$ appears in $\M_n$ independently with probability $q_n$;
\item each monomial of the form $X_i X_j$ with $i\ne j$ appears independently with probability $p_n$.
\end{myitemize}

The goal is to estimate in which cases the random variable $\nu(\M_n)$ grows sufficiently
quickly so that the assumptions of Theorem \ref{thm:semiregfriendly}
are satisfied asymptotically with large probability.
In order to estimate $\nu(\M_n)$, we first forget the meaning of the graph in terms of monomials and count the number of isolated edges in a random graph $G$ in this variant of the Erd\"os-Renyi model.

\begin{proposition}\label{prop:probaNi}
Let $G$ be a random simple graph on $n+1$ vertices. Each vertex has a loop with probability $q\in [0,1]$ and an edge between any two vertices appear with probability $p\in [0,1]$. Let $G'$ be the subgraph obtained by restricting $G$ to the vertices with a loop and $\Ni$ be the random variable counting the number of isolated edges in $G'$. Then $\Ni$ has expected value and variance 
\begin{align*}
\EV(\Ni)=&\displaystyle\binom{n+1}{2}\,q^2\,p\,(1-q(1-(1-p)^2))^{n-1},  \\
\Var(\Ni)=&\displaystyle\EV(\Ni)-\EV(\Ni)^2+\\&6\,\binom{n+1}{4}\,q^4\,p^2\,(1-p)^4(1-q(1-(1-p)^4))^{n-3}.\numberthis \label{eq:var}
\end{align*}
\end{proposition}

\begin{proof}
  For each possible edge $e$ between two vertices $i\neq j$, we denote by $X_e$
  the random variable taking the value $1$ if $e$ is an isolated edge of $G'$,
  and $0$ otherwise. The probability that $i$ and $j$ appear as vertices with
  loops in $G$ is $q^2$. Hence, the probability that the edge $e$ lies in $G'$ is $q^2p$. Moreover, for a given vertex $k\neq i,j$, the probability that $k$ appears in $G'$ and at least one of the edges $(i,k)$ and $(j,k)$ belong to $G$ is $q(1-(1-p)^2)$. There are $n-1$ other vertices than $i$ and $j$ in $G$, hence $X_e$ follows a Bernoulli law of parameter $q^2p(1-q(1-(1-p)^2))^{n-1}$. It follows that
\[
  \EV(\Ni)  =  \sum_e \EV(X_e) 
 =  \binom{n+1}{2}q^2p(1-q(1-(1-p)^2))^{n-1} \] 

The computation of the variance can be done similarly.
\end{proof}

We now apply the previous proposition in the case where $p$ and $q$ depends on $n$, and analyze the convergence of $\EV(\Ni)$ and $\Var(\Ni)$ as $n$ grows to infinity.

\begin{corollary}\label{coro:EVvarNi}
  Let $p_n=\Theta(n^{-1})$ and $q_n=\Theta(n^\beta)$. With the notation of Proposition~\ref{prop:probaNi}, if $-1/2<\beta<0$ then $\EV(\Ni)=\Theta\left(n^{2\beta+1}\right)$ and $\Var(\Ni)=\Theta\left(n^{2\beta+1}\right)$.
\end{corollary}

\begin{proof}
First, note that $ \log\pac{(1-q_n(1-(1-p_n)^2))^{n-1}} =
-2 n p_nq_n+O\left(n^{-1}\right) $ since $\beta<0$. This shows that
$\EV(\Ni) = \frac{q_n^2 p_nn^2}{2}e^{-2 n p_nq_n}(1+O(n^{-1}))$.
The claim on the asymptotic behavior of $\EV(\Ni)$ follows from
$e^{-2 n p_nq_n}=\Theta(1)$.  Next, let $\lambda$ denote the last
summand in Eq. \eqref{eq:var}, namely
$\lambda=\Var(\Ni)-\EV(\Ni)+\EV(\Ni)^2$.  The asymptotic behavior of
$\lambda$ can be obtained by a similar analysis:
\[
\log\pac{(1-q_n(1-(1-p_n)^4))^{n-3}} = -4 n p_nq_n+O\left( n^{-1}\right),
\]
hence $\lambda=\frac{q_n^4 p_n^2 n^4}{4}e^{-4 n p_nq_n} (1+O(n^{-1}))$. Notice that $\EV(\Ni)^2 = \frac{q_n^4 p_n^2 n^4}{4} e^{-4 n p_nq_n}(1+O(n^{-1}))$. Consequently, $\EV(\Ni)^2-\lambda=O(n^{4\beta+1})$, since $e^{-4 n p_nq_n}=\Theta(1)$. Finally, putting all the estimates together, we obtain $\Var(\Ni)=\Theta(n^{2\beta+1})+O(n^{4\beta+1})=\Theta(n^{2\beta+1})$ since $\beta<0$.
 \end{proof}

Finally, we relate the distribution of $\Ni$ with the probability that the assumptions of Theorem \ref{thm:semiregfriendly} hold for fewnomial systems with $\lvert\M\rvert=n+k+1$. If one wants that $\EV(\lvert\M_n\rvert)=n+k+1$ for some fixed $k$ and that the expected number of squares is $(n+1)^{1/2+\varepsilon}$, then one has to choose $q_n=(n+1)^{-1/2+\varepsilon}$ and $p_n=(n+k+1-(n+1)q_n)/\binom{n+1}{2}$. The asymptotic expected behavior of the matching number in that case is described by the following statement:

\begin{lemma}\label{lem:mainProbaResult}
Let $\M_n$ be a sequence of random supports where each square monomial appears with probability $q_n$, and each square-free monomial appears with probability $p_n$. If $p_n=\Theta(n^{-1})$ and $q_n=\Omega(n^{-1/2+\varepsilon})$, with $0<\varepsilon<1/2$, then for any $\ell\in\N$, 
$\Prob\left(\nu(\M_n)\geq \ell\right)$
 tends towards $1$ as $n$ grows.
\end{lemma}

\begin{proof}
Chebyshev's inequality implies that
\begin{align*}
\Prob(\Ni\leq\EV(\Ni)/2)\leq& \Prob(\mid\Ni-\EV(\Ni)\mid\geq \EV(\Ni)/2)
\leq \dfrac{4\Var(\Ni)}{\EV(\Ni)^2}\\
= &O\left(n^{-2(-1/2+\varepsilon)-1}\right)
= O\left(n^{-2 \varepsilon}\right).
\end{align*}
Next, notice that $\EV(\Ni)/2=\Theta\left(n^{2\varepsilon}\right)$ by Corollary \ref{coro:EVvarNi}. Also, note that $\Ni\leq\nu(\M_n)$, so that for $n$ sufficiently large, we have 
$\displaystyle\Prob\left(\nu(\M_n)\leq n^\varepsilon\right)\leq\displaystyle\Prob\left(\Ni\leq n^{2\varepsilon}\right)= O\left(n^{-2 \varepsilon}\right),$
which tends towards $0$ as $n$ grows.
\end{proof}

Next, we show that these estimates also hold for a different model of random monomial supports. For $n\in\N$ and two integers $a,b\in\N$, we consider the random sets $\U_{n,a,b}$ of quadratic monomials in $\K[X_0,\ldots, X_n]$ distributed uniformly at random among those that contain $a$ non-squares and $b$ squares.

\begin{theorem}\label{thm:uniformmodel}
  Let $k$ be a fixed integer, $a_n,b_n\in\N$ be such
  $a_n+b_n=n+k+1$, and $\U_{n,a_n,b_n}$ be a subset of quadratic
  monomials in $\K[X_0,\ldots, X_n]$ distributed uniformly at random
  among those that contain $a_n$ non-square monomials and $b_n$ squares. Assume further that $b_n=\Omega(n^{1/2+\varepsilon})$, for $\varepsilon>0$. 
Then the probability that the assumptions of Theorem \ref{thm:semiregfriendly}
with $m=n$ are satisfied for $\U_{n,a_n,b_n}$ tends towards $1$ as $n$ grows. 
\end{theorem}

\begin{proof}
The proof of this theorem is technical and is similar to the classical
techniques to prove properties of random graphs in the Erd\"os-Renyi models
\cite{erdos1960evolution}. Details are provided in the appendix.
\end{proof}

Theorem \ref{thm:introrandgraph} is a direct consequence of Theorem
\ref{thm:uniformmodel} and is obtained by dehomogenization.

\mysection{Systems with all the squares}

\label{sec:allsquares}

Next, we investigate the special case of fewnomial systems
where all the squares $X_i^2$ belong to
$\M$. This corresponds to a limit case of Theorem \ref{thm:uniformmodel}: $\varepsilon=1/2$. In this setting, the Newton polytopes of the
polynomials are the same as those of dense quadratic polynomials, hence these systems have generically $2^n$ solutions in $\Kclos^n$. In the sequel of this section, $\M$ is a set of monomials of degree at most $2$ in $\K[X_1,\ldots,X_n]$, of cardinality $n+k+1$, and which contains the constant $1$ and all the squares $X_i^2$. We also assume that $n>2k$.

We let $\ell$ denote the number of variables $X_i$ which appear in a square-free monomial in $\M$. Hence $\ell\leq 2 k$. For a $0$-dimensional system $(f_1,\ldots, f_n)\in\polM^n$, we let $S$ denote the $n\times(n-\ell)$ matrix which contains the coefficients of the squares $X_i^2$ such that $X_i$ does not appear in a square-free monomial in $\M$.

\begin{proposition}\label{prop:orbit}
Let $(f_1,\ldots, f_n)\in\polM^n$ be a $0$-dimensional system with support $\M$. Then the system $f_1=\dots=f_n=0$ has at most $2^n$ solutions in $\Kclos^n$. If the matrix $S$ has full rank, then the solutions are the orbits of at most $2^{\ell}$ points under the action of $\left(\Z/2\Z\right)^{n-\ell}$ given by 
$$\begin{array}{rccl}
\chi:&\left(\Z/2\Z\right)^{n-\ell}\times \Kclos^n&\rightarrow&\quad\Kclos^n\\
&(\mathbf e_i,(a_1,\dots,a_n))&\mapsto& (a_1,\dots,-a_{i_j},\ldots,a_n)
\end{array},$$
where the set $\{i_j\}$ is the set of indices such that $X_{i_j}$ does not appear in a square-free monomial in $\M$.
\end{proposition}

\begin{proof}
Up to a permutation of the indices, we can assume w.l.o.g. that $X_1,\ldots, X_{n-\ell}$ are the variables that does not appear in a square-free monomial in $\M$.
Since the matrix $S$ is full-rank, we perform Gaussian elimination to remove the squares $X_{i_j}^2$ which do not belong to an edge of the graph. This provides us with an equivalent system of the form
$$
\left\{\begin{array}{l}
X_1^2-g_1(X_{n-\ell+1},\ldots,X_n) = 0\\
\quad\quad\quad\vdots\\
X_{n-\ell}^2-g_{n-\ell}(X_{n-\ell+1},\ldots,X_n) = 0\\
h_1(X_{n-\ell+1},\ldots,X_n) = \dots =
h_\ell(X_{n-\ell+1},\ldots,X_n)=0.
\end{array}\right.
$$ We end up with a system $(h_1,\ldots, h_\ell)$ of dense homogeneous
polynomials in $\ell$ variables. Note that $\ell$ is bounded by $2k$,
which does not depend on $n$. Consequently, this system can be solved
within a constant number of operations as $n$ grows. By B\'ezout
theorem, this system has at most $2^{\ell}$ solutions.  Finally, if
$(a_{n-\ell+1},\ldots, a_n)$ is a solution of $h_1=\dots=h_\ell=0$,
then $(\pm \sqrt{g_1(a_{n-\ell+1},\ldots, a_n)},\ldots,\linebreak \pm
\sqrt{g_{n-\ell}(a_{n-\ell+1},\ldots, a_n)},a_{n-\ell+1},\ldots,
a_n)$ is a solution of the input system. Moreover, all solutions are
of this form.
\end{proof}

Therefore, even though the number of solutions of such systems depends
exponentially on $n$, they can be conveniently represented. Moreover, we show next that computing this representation can be achieved within a number of operations in $\K$ which is polynomial in~$n$:

\begin{corollary}
  Let $(f_1,\ldots, f_n)\in\polM^n$ be polynomials with support $\M$
  satisfying the above assumptions ($\lvert\M\rvert=n+k+1$, all
  squares are in $\M$, $S$ has full-rank) and $\mu$ be a square-free
  monomial. For fixed $k$, Problem \ref{pb:solving} with input $(f_1,\ldots,
  f_n)$ and $\mu$ can be solved within
  $O\left(n^\omega\right)$ arithmetic operations as $n$ grows, where $\omega$ is a feasible
  exponent for matrix multiplication.
\end{corollary}
\begin{proof}
  With the same notations as in the proof of Proposition
  \ref{prop:orbit}, and by noticing that $\langle h_1,\ldots,
  h_\ell\rangle\cap \K[\mu]=\langle f_1,\ldots, f_n\rangle\cap
  \K[\mu]$, solving Problem \ref{pb:solving} with input $(h_1,\ldots,
  h_\ell)$ and $\mu$ yields a solution to Problem \ref{pb:solving}
  with input $(f_1,\ldots, f_n)$ and $\mu$. Solving
  Problem \ref{pb:solving} with input $h_1,\ldots, h_\ell$ can be achieved
  within a time complexity which does not depend on $n$. Consequently, the
  only complexity that depends on $n$ is the cost of computing the
  polynomials $h_1,\ldots, h_\ell$. This is done by linear algebra,
  within $O\left(n^\omega\right)$ operations in $\K$.
\end{proof}


\mysection{Experimental results}\label{sec:experiments}

In this section, we describe experimental results, validating the
theoretical results and illustrating their practical relevance. In particular, our prototype implementation of the algorithm in the proof of Corollary~\ref{cor:complNullstellensatz} is able to compute
Nullstellensatz' certificates of inconsistency for systems of 30000
equations and 30000 unknowns generated from the uniform model in Theorem \ref{thm:uniformmodel}.
This may be compared to the practical timings for solving the same
problem with dense generic quadratic systems, where 20 unknowns is
already a difficult challenge due to the exponential size of the
certificates.

\smallskip

{\bf Experimental setting.} $\K$ is the finite field ${\rm GF}(65521)$. The experimental procedure depends on parameters $n,k$ and $\beta$:
\begin{myitemize}
\item generate a random support $\M$ of $n+k+1$ monomials of degree at most $2$, containing $1$ and $\lfloor n^\beta\rfloor$ squares. The subsets of square monomials and non-square monomials are respectively chosen uniformly at random;
\item generate a random system of $n$ equations with support $\M$, where all the coefficients are chosen uniformly at random in $\K$;
\item return ``success'' if our implementation returns a relation
$1=\sum_{i=1}^n h_i f_i,$
with $h_i\in \polM$, else return ``failure''.
\end{myitemize}
By Theorem \ref{thm:uniformmodel}, for any choice of parameters $k\in\N$ and
$0.5<\beta<1$, the probability that ``success'' is returned should
tend towards $1$ as $n$ grows. 

\smallskip

\begin{figure}\centering
\begin{tikzpicture}[scale=0.8]
\begin{axis}[xmin=2500,xmax=30000,ymin=0,ymax=100,
xlabel style={at={(0.93,0.1)},anchor=south}, 
ylabel style={at={(0.05,1.09)},anchor=east}, 
xlabel={\(n\)},
ylabel={$\%$ of success},
cycle list={%
{blue},
{red},
{,
mark options={fill=brown!40},mark=otimes*}},
legend style={at={(1.02,0.5)},anchor=west}]

%
%
%

\addplot[blue,thick,mark=*] coordinates {
(30000,99.41)
(20000,99.31)
(10000,98.82)
(5000,97.75)
(2500,95.70)
};

\addplot[thick,mark=*] coordinates {
(30000,99.41)
(20000,98.73)
(10000,96.67)
(5000,94.92)
(2500,93.06)
};

\addplot[thick,mark=*] coordinates {
(30000,96.77)
(20000,95.89)
(10000,92.48)
(5000,91.11)
(2500,87.79)
};

\addplot[thick,mark=*] coordinates {
(30000,94.53)
(20000,91.69)
(10000,91.40)
(5000,87.98)
(2500,85.93)
};

\addplot[thick,mark=*] coordinates {
(30000,91.30)
(20000,90.52)
(10000,87.98)
(5000,84.17)
(2500,79.49)
};

\addplot[thick,mark=*] coordinates {
(40000,84.76)
(30000,86.13)
(20000,83.39)
(10000,81.83)
(5000,79.39)
(2500,75.68)
};

\addplot[thick,mark=*] coordinates {
(30000,80.27)
(20000,77.24)
(10000,74.51)
(5000,72.55)
(2500,71.38)
};

\addplot[thick,mark=*] coordinates {
(30000,71.77)
(20000,73.63)
(10000,69.04)
(5000,68.94)
(2500,64.94)
};

\addplot[thick,mark=*] coordinates {
(30000,66.79)
(20000,66.11)
(10000,65.03)
(5000,60.93)
(2500,62.10)
};

\addplot[orange,thick,mark=*] coordinates {
(30000,58.98)
(20000,62.98)
(10000,59.27)
(5000,62.30)
(2500,58.59)
};

\addplot[red,thick,mark=*] coordinates {
(30000,55.95)
(20000,58.49)
(10000,54.58)
(5000,55.95)
(2500,56.64)
};

\addplot[yellow,thick,mark=*] coordinates {
(30000,51.17)
(20000,52.83)
(10000,52.44)
(5000,52.92)
(2500,52.44)
};

\addplot[thick,mark=*] coordinates {
(30000,48.33)
(20000,47.75)
(10000,46.28)
(5000,48.63)
(2500,47.94)
};

\addplot[thick,mark=*] coordinates {
(30000,43.06)
(20000,47.26)
(10000,47.75)
(5000,47.46)
(2500,45.60)
};

\addplot[blue,thick,mark=*] coordinates {
(30000,45.70)
(20000,41.11)
(10000,41.11)
(5000,40.23)
(2500,44.53)
};


\legend{{$\beta=0.6$},{$\beta=0.59$},{$\beta=0.58$},{$\beta=0.57$},{$\beta=0.56$},{$\beta=0.55$},{$\beta=0.54$},{$\beta=0.53$},{$\beta=0.52$},{$\beta=0.51$},{$\beta=0.5$},{$\beta=0.49$},{$\beta=0.48$},{$\beta=0.47$},{$\beta=0.45$}}
\end{axis}
\end{tikzpicture}

\caption{$k=2$ fixed, $n$ grows, several value of $\beta$. Every point is an
  average over 1000 tests. The relative positions of the curves follow the
  values of $\beta$.\label{fig:betan}}
\end{figure}
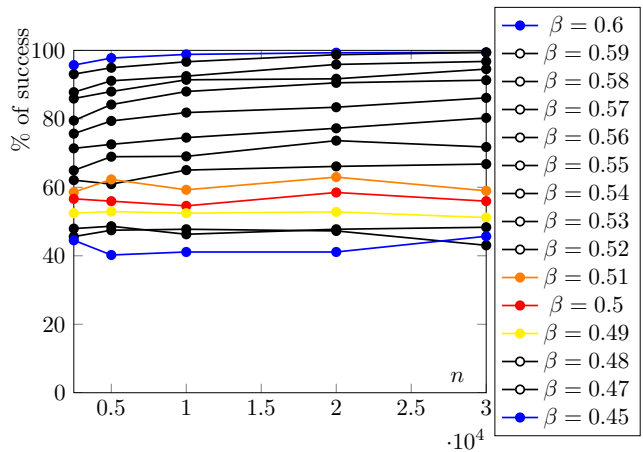

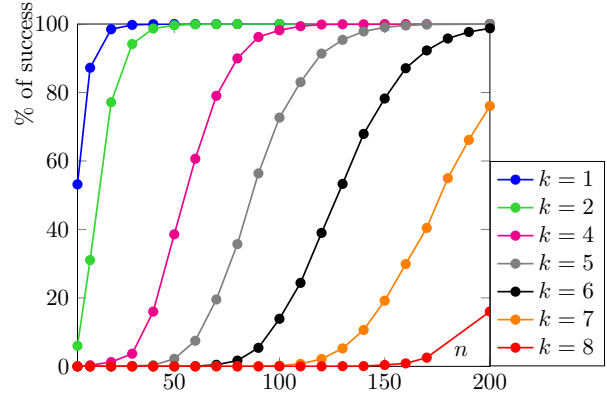
\begin{figure}\centering
\begin{tikzpicture}[thick, scale=0.8]
\begin{axis}[xmin=4,xmax=200,ymin=0,ymax=100,
xlabel style={at={(0.93,0.1)},anchor=south}, 
ylabel style={at={(0.05,1.09)},anchor=east}, 
xlabel={\(n\)},
ylabel={$\%$ of success},
cycle list={%
{blue},
{blue,dashed},
{only marks,red,mark=square,thick},
{blue},
{red,dashed},
{,
mark options={fill=brown!40},
mark=otimes*}},
legend style={at={(1,0.3)},anchor=west}]
\addplot[blue,thick,mark=*] coordinates {
(4,53.17)
(10,87.22)
(20,98.47)
(30,99.73)
(40,99.95)
(50,99.99)
(60,100.00)
(70,100.00)
(80,100.00)
(100,100.00)
(200,99.99)
};
\addplot[green!80!black!80,thick,mark=*] coordinates {
(4,6.01)
(10,31.04)
(20,77.16)
(30,94.16)
(40,98.68)
(50,99.62)
(60,99.91)
(70,99.97)
(80,100.00)
(100,99.99)
(200,99.99)
};
\addplot[magenta,thick,mark=*] coordinates {
(4,0)
(10,.33)
(20,1.25)
(30,3.70)
(40,16.00)
(50,38.57)
(60,60.64)
(70,79.02)
(80,89.96)
(90,96.21)
(100,98.18)
(110,99.36)
(120,99.85)
(130,99.93)
(140,99.92)
(150,99.98)
(160,99.98)
(170,99.98)
(200,99.98)
};
\addplot[gray,thick,mark=*] coordinates {
(4,0)
(10,.03)
(20,.04)
(30,.15)
(40,.35)
(50,2.19)
(60,7.48)
(70,19.54)
(80,35.71)
(90,56.37)
(100,72.67)
(110,83.03)
(120,91.37)
(130,95.38)
(140,97.83)
(150,99.05)
(160,99.57)
(170,99.86)
(200,99.97)
};
\addplot[black,thick,mark=*] coordinates {
(4,0)
(10,0)
(20,0)
(30,0)
(40,0)
(50,.01)
(60,.06)
(70,.46)
(80,1.68)
(90,5.42)
(100,13.90)
(110,24.39)
(120,38.99)
(130,53.30)
(140,67.90)
(150,78.24)
(160,87.09)
(170,92.28)
(180,95.77)
(190,97.68)
(200,98.75)
};
\addplot[orange,thick,mark=*] coordinates {
(4,0)
(10,0)
(20,0)
(30,0)
(40,0)
(50,0)
(60,0)
(70,0)
(80,0)
(90,.03)
(100,.24)
(110,.73)
(120,2.16)
(130,5.25)
(140,10.63)
(150,19.20)
(160,29.88)
(170,40.45)
(180,54.96)
(190,66.15)
(200,76.06)
};
\addplot[red,thick,mark=*] coordinates {
(4,0)
(10,0)
(20,0)
(30,0)
(40,0)
(50,0)
(60,0)
(70,0)
(80,0)
(90,0)
(100,0)
(110,0)
(120,.01)
(130,.06)
(140,.08)
(150,.41)
(160,.87)
(170,2.55)
(200,16.02)
};
\legend{{$k=1$}, {$k=2$},  {$k=4$} , {$k=5$} , {$k=6$} , {$k=7$},{$k=8$} }
\end{axis}
\end{tikzpicture}

\caption{$\beta=0.9$ fixed, $n$ grows, several value of $k$. Every point is an
  average over 10000 tests.\label{fig:kn}}
\end{figure}

First, we study the dependence of the asymptotic behavior on the choice of
$\beta$. To this end, we fix $k=1$ and we look at the
experimental probability of success as $n$ grows. Experimental results
are reported in Figure~\ref{fig:betan}. The results are in accordance
with Theorem~\ref{thm:uniformmodel}: when $\beta > 0.5$, the probability that such systems
have no solution and that there exists a Nullstellensatz certificate in $\polM$ seems to tend to $1$ as $n$ grows. We also observe that the
convergence seems to depend strongly on $\beta$: when $\beta$ becomes
close to the limit value $0.5$, the speed of convergence seems to
decrease.
Next, we focus on the dependency on $k$. We fix $\beta=0.9$ and let
$n$ grow for different values of $k$.  Experiments are reported in
Figure \ref{fig:kn}. Finally, we look at quadratic supports $\M$ of cardinality $n+k+1$
generated uniformly at random without any constraint on the number of
squares. This case is not covered by the analysis of this paper and
experiments show a different behavior: the probability of success of the algorithm
does not seem to tend to $1$ as $n$ grows, contrary to the case
$\beta>0.5$. However, this probability seems to converges to a nonzero
value.

\begin{conjecture}
  Let $k\in\N$ be a fixed integer. For $n\in\N$, let $\M_n$ be a random
  subset of monomials in $n$ variables of degree at most $2$, uniformly distributed
  among those of cardinality $n+k+1$ that contain~$1$. Let $f_1,\ldots, f_n\in\polM$ be a
  system with support $\M$ and generic coefficients. Then the
  probability that there exist $h_1,\ldots, h_n\in\polM$ such that
  $\sum_{i=1}^n f_i h_i = 1$ tends to a nonzero value as $n$ grows.
\end{conjecture}

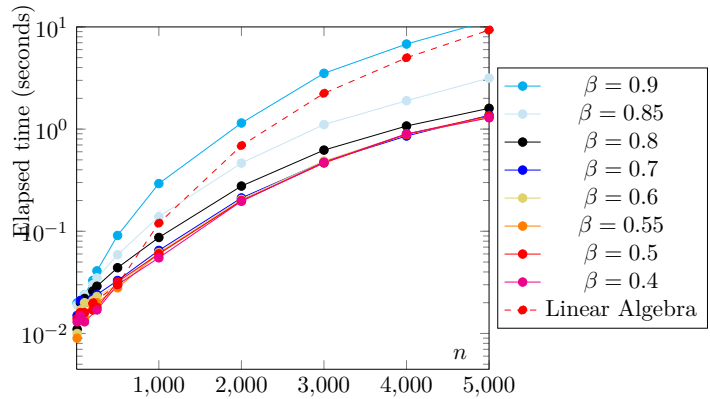
\begin{figure}\centering
\begin{tikzpicture}[scale=0.8]
\begin{axis}[xmin=4,xmax=5000,ymin=0,ymax=10,
xlabel style={at={(0.93,0.1)},anchor=south}, 
ylabel style={at={(0.05,1.09)},anchor=east}, 
xlabel={\(n\)},
ylabel={Elapsed time (seconds)},
ymode=log,
cycle list={%
{blue},
{blue,dashed},
{only marks,red,mark=square,thick},
{blue},
{red,dashed},
{,
mark options={fill=brown!40},
mark=otimes*}},
legend style={at={(1.02,0.5)},anchor=west}]

\addplot[cyan,mark=*] coordinates {
(10,0.020)(50,0.021)(100,0.024)(200,0.033)(250,0.041)(500,0.091)(1000,0.293)(2000,1.147)(3000,3.516)(4000,6.798)(5000,11.489)
};
\addplot[cyan!80!black!20,mark=*] coordinates {
(10,0.019)(50,0.021)(100,0.024)(200,0.030)(250,0.035)(500,0.059)(1000,0.139)(2000,0.465)(3000,1.107)(4000,1.896)(5000,3.155)
};
\addplot[black,mark=*] coordinates {
(10,0.011)(50,0.020)(100,0.022)(200,0.026)(250,0.029)(500,0.044)(1000,0.087)(2000,0.277)(3000,0.623)(4000,1.072)(5000,1.597)
};
\addplot[blue,mark=*] coordinates {
(10,0.015)(50,0.021)(100,0.020)(200,0.022)(250,0.024)(500,0.033)(1000,0.065)(2000,0.212)(3000,0.479)(4000,0.856)(5000,1.366)
};
\addplot[yellow!80!black!80,mark=*] coordinates {
(10,0.010)(50,0.017)(100,0.020)(200,0.021)(250,0.023)(500,0.030)(1000,0.062)(2000,0.205)(3000,0.487)(4000,0.893)(5000,1.331)
};
\addplot[orange,mark=*] coordinates {
(10,0.009)(50,0.016)(100,0.014)(200,0.019)(250,0.020)(500,0.028)(1000,0.061)(2000,0.197)(3000,0.469)(4000,0.885)(5000,1.295)
};
\addplot[red,mark=*] coordinates {
(10,0.014)(50,0.016)(100,0.016)(200,0.017)(250,0.018)(500,0.032)(1000,0.060)(2000,0.200)(3000,0.466)(4000,0.895)(5000,1.343)
};
\addplot[magenta,mark=*] coordinates {
(10,0.013)(50,0.015)(100,0.013)(250,0.017)(500,0.030)(1000,0.055)(2000,0.197)(3000,0.474)(4000,0.898)(5000,1.290)
};

\addplot[red,dashed,mark=*] coordinates {
(10,0.000)(50,0.000)(100,0.000)(200,0.020)(250,0.000)(500,0.030)(1000,0.120)(2000,0.690)(3000,2.240)(4000,4.990)(5000,9.310)
};

\legend{{$\beta=0.9$},{$\beta=0.85$},{$\beta=0.8$},{$\beta=0.7$},{$\beta=0.6$},{$\beta=0.55$},{$\beta=0.5$},{$\beta=0.4$},{Linear Algebra}}

\end{axis}
\end{tikzpicture}
  \caption{Timings for the computation of the certificate of
    inconsistency, $k=2$.\label{fig:timings}}
\end{figure}

Finally, we report in Figure \ref{fig:timings} on experiments about the
efficiency our prototype implementation for
computing Nullstellensatz certificates. The experiments were conducted on a
Mac Retina 2.8Ghz Intel Core i7, and linear algebra computations were performed
with {\tt Magma V2.20-3}.
We see in these experiments that
systems with several thousands of variables can be handled in a few seconds.
The algorithm works in two steps: first we reduce the quadratic system with
linear algebra (the complexity of this step is independent of $\beta$ and is
represented by the dashed curve); then, the matrix in degree $4$ (multiplying
all the reduced polynomials by all the monomials in $\M$) is constructed and
reduced. The time of this second step depends on $\beta$ and is indicated by the
plain curves. Therefore, these graphs seem to indicate that the cost of computing
certificates of inconsistency for these systems is approximately twice the time
of computing the row echelon form of a dense $n\times n$ matrix.

\bibliographystyle{abbrv}
\bibliography{biblio}

\begin{thebibliography}{10}

\bibitem{beame1994lower}
P.~Beame, R.~Impagliazzo, J.~Kraj{\'\i}{\v{c}}ek, T.~Pitassi, and
  P.~Pudl{\'a}k.
\newblock Lower bounds on {H}ilbert's {N}ullstellensatz and propositional
  proofs.
\newblock In {\em Foundations of Computer Science, 1994 Proceedings., 35th
  Annual Symposium on}, pages 794--806. IEEE, 1994.

\bibitem{bernshtein1975number}
D.~Bernshtein.
\newblock The number of roots of a system of equations.
\newblock {\em Functional Analysis and its Applications}, 9(3):183--185, 1975.

\bibitem{bertrand2006polynomial}
B.~Bertrand, F.~Bihan, and F.~Sottile.
\newblock Polynomial systems with few real zeroes.
\newblock {\em Mathematische Zeitschrift}, 253(2):361--385, 2006.

\bibitem{bihan2007new}
F.~Bihan and F.~Sottile.
\newblock New fewnomial upper bounds from {G}ale dual polynomial systems.
\newblock {\em Moscow mathematical journal}, 7(3):387--407, 2007.

\bibitem{brownawell1987bounds}
W.~D. Brownawell.
\newblock Bounds for the degrees in the {N}ullstellensatz.
\newblock {\em Annals of Mathematics}, pages 577--591, 1987.

\bibitem{canny2000subdivision}
J.~Canny and I.~Emiris.
\newblock A subdivision-based algorithm for the sparse resultant.
\newblock {\em Journal of the ACM}, 47(3):417--451, 2000.

\bibitem{DBLP:journals/corr/CifuentesP14}
D.~Cifuentes and P.~Parrilo.
\newblock Exploiting chordal structure in polynomial ideals: a {G}r{\"{o}}bner
  bases approach.
\newblock {\em arXiv}, abs/1411.1745, 2014.

\bibitem{clegg1996using}
M.~Clegg, J.~Edmonds, and R.~Impagliazzo.
\newblock Using the {G}roebner basis algorithm to find proofs of
  unsatisfiability.
\newblock In {\em Proceedings of the twenty-eighth annual ACM symposium on
  Theory of computing}, pages 174--183. ACM, 1996.

\bibitem{CoxLitSch11}
D.~A. Cox, J.~B. Little, and H.~K. Schenck.
\newblock {\em Toric varieties}.
\newblock AMS, 2011.

\bibitem{crupi2011cohen}
M.~Crupi, G.~Rinaldo, and N.~Terai.
\newblock Cohen-macaulay edge ideal whose height is half of the number of
  vertices.
\newblock {\em Nagoya Mathematical Journal}, 201:117--131, 2011.

\bibitem{edmonds1965paths}
J.~Edmonds.
\newblock Paths, trees, and flowers.
\newblock {\em Canadian Journal of mathematics}, 17(3):449--467, 1965.

\bibitem{erdos1960evolution}
P.~Erd{\"o}s and A.~R{\'e}nyi.
\newblock On the evolution of random graphs.
\newblock {\em Publ. Math. Inst. Hung. Acad. Sci}, 5:17--61, 1960.

\bibitem{FauSpaSva14}
J.-C. Faug{\`e}re, P.-J. Spaenlehauer, and J.~Svartz.
\newblock Sparse {G}r{\"o}bner bases: the unmixed case.
\newblock In {\em Proceedings of ISSAC 2014}, 2014.

\bibitem{Fit}
N.~Fitchas and A.~Galligo.
\newblock Nullstellensatz effectif et conjecture de {S}erre (th\'eor\`eme de
  {Q}uillen-{S}uslin) pour le calcul formel.
\newblock {\em Mathematische Nachrichten}, 149(1):231--253, 1990.

\bibitem{froberg1990stanley}
R.~Fr{\"o}berg.
\newblock On {S}tanley-{R}eisner rings.
\newblock {\em Banach Center Publications}, 26(2):57--70, 1990.

\bibitem{herzog2005distributive}
J.~Herzog and T.~Hibi.
\newblock Distributive lattices, bipartite graphs and alexander duality.
\newblock {\em Journal of Algebraic Combinatorics}, 22(3):289--302, 2005.

\bibitem{herzog2003monomial}
J.~Herzog, T.~Hibi, and X.~Zheng.
\newblock Monomial ideals whose powers have a linear resolution.
\newblock {\em Mathematica Scandinavica}, 95(1):23--32, 2004.

\bibitem{huber1995polyhedral}
B.~Huber and B.~Sturmfels.
\newblock A polyhedral method for solving sparse polynomial systems.
\newblock {\em Mathematics of Computation}, 64(212):1541--1555, 1995.

\bibitem{khovanskii1980class}
A.~Khovanskii.
\newblock On a class of systems of transcendental equations.
\newblock {\em Soviet Mathematics Doklady}, 22(3):762--765, 1980.

\bibitem{DBLP:journals/corr/KoiranPT13}
P.~Koiran, N.~Portier, and S.~Tavenas.
\newblock On the intersection of a sparse curve and a low-degree curve: A
  polynomial version of the lost theorem.
\newblock {\em Discrete and Computational Geometry}, 53(1):48--63, 2015.

\bibitem{DBLP:journals/corr/KoiranPTT13}
P.~Koiran, N.~Portier, S.~Tavenas, and S.~Thomass\'e.
\newblock A $\tau$-conjecture for newton polygons.
\newblock {\em Foundations of Computational Mathematics}, pages 1--13, 2014.

\bibitem{kollar1988sharp}
J.~Koll{\'a}r.
\newblock Sharp effective {N}ullstellensatz.
\newblock {\em Journal of the American Mathematical Society}, 1(4):963--975,
  1988.

\bibitem{krick2001sharp}
T.~Krick, L.~M. Pardo, M.~Sombra, et~al.
\newblock Sharp estimates for the arithmetic {N}ullstellensatz.
\newblock {\em Duke Mathematical Journal}, 109(3):521--598, 2001.

\bibitem{kushnirenko1976newton}
A.~G. Kushnirenko.
\newblock Newton polytopes and the {B}ezout theorem.
\newblock {\em Functional Analysis and its Applications}, 10(3):233--235, 1976.

\bibitem{Laz77}
D.~Lazard.
\newblock Alg\`ebre lin\'eaire sur $k[x_1,\ldots,x_n]$ et \'elimination.
\newblock {\em Bull. Soc. Math. France}, 105:165--190, 1977.

\bibitem{Leg14}
F.~{Le Gall}.
\newblock Powers of tensors and fast matrix multiplication.
\newblock In {\em Proceedings of ISSAC'14}, 2014.

\bibitem{lovasz1986matching}
L.~Lov{\'a}sz and M.~D. Plummer.
\newblock {\em Matching theory}.
\newblock AMS, 1986.

\bibitem{MilStu05}
E.~Miller and B.~Sturmfels.
\newblock {\em Combinatorial commutative algebra}, volume 227.
\newblock Springer Verlag, 2005.

\bibitem{semaev2008solving}
I.~Semaev.
\newblock On solving sparse algebraic equations over finite fields.
\newblock {\em Designs, Codes and Crypto.}, 49(1-3):47--60, 2008.

\bibitem{shub1996intractability}
M.~Shub and S.~Smale.
\newblock On the intractability of {H}ilbert's {N}ullstellensatz and an
  algebraic version of {"P= NP"}.
\newblock {\em Duke Mathematical Journal}, 81(1):47--54, 1996.

\bibitem{sombra1999sparse}
M.~Sombra.
\newblock A sparse effective {N}ullstellensatz.
\newblock {\em Advances in Applied Mathematics}, 22(2):271--295, 1999.

\bibitem{storjohann2000algorithms}
A.~Storjohann.
\newblock Algorithms for matrix canonical forms.
\newblock {\em Ph.D. thesis}, 2000.

\bibitem{Stu91}
B.~Sturmfels.
\newblock Sparse elimination theory.
\newblock In {\em Proceedings of Computational Algebraic Geometry and
  Commutative Algebra}, pages 377--396. Cambridge Univ. Press, 1991.

\bibitem{Stu96}
B.~Sturmfels.
\newblock {\em Gr{\"o}bner bases and convex polytopes}.
\newblock AMS, 1996.

\bibitem{verschelde1994homotopies}
J.~Verschelde, P.~Verlinden, and R.~Cools.
\newblock Homotopies exploiting {N}ewton polytopes for solving sparse
  polynomial systems.
\newblock {\em SIAM Journal on Numerical Analysis}, 31(3):915--930, 1994.

\bibitem{Woo10}
R.~Woodroofe.
\newblock Matchings, coverings, and {C}astelnuovo-{M}umford regularity.
\newblock {\em J. of Comm. Algebra}, 2:287--304, 2014.

\end{thebibliography}

\newpage

\appendix

\section{Proof of Theorem 4.4}

  Set $$p_n=a_n/\binom{n+1}{2} \text{ and }q_n=b_n/(n+1),$$ and let $\M_n$ be
  the random support constructed as above with respect to the
  probabilities $p_n$ and $q_n$. We let $\MS_n$ denote
  the subset of squares in $\M_n$ and
  $\MNS_n$ denote the subset of nonsquare monomials
  in $\M_n$. Also, we set $\ell=(k^2+3k+2)/2$.
  First, we notice that $\Prob(\nu(\M_n)\geq
    \ell)$ equals

  \begin{eqnarray}&\displaystyle\sum_{\substack{0\leq i\leq\binom{n+1}2\\0\leq j\leq
  n}}\Prob(\nu(\M_n)\geq \ell \mid \lvert\MS\rvert=i\text{ and
  }\lvert\MNS\rvert=j)\,c_{n,i,j} \nonumber\\
  &=\displaystyle\sum_{\substack{0\leq i\leq\binom{n+1}2\\0\leq j\leq
n}}\Prob(\nu(\U_{n,i,j})\geq\ell)\,c_{n,i,j},\label{eq:cij}
\end{eqnarray}
where
 $c_{n,i,j}=
p_n^i(1-p_n)^{\binom{n+1}2-i}q_n^j(1-q_n)^{n+1-j}\binom{n+1}j\binom{\binom{n+1}2}{i}$
is the probability that $\lvert\MS_n\rvert=i$ and $\lvert\MNS_n\rvert=j$.
Since the matching number is monotone with respect to the subgraph ordering, we obtain
$$i_1\geq i_2 \text{ and } j_1\geq j_2 \Longrightarrow \Prob(\nu(\U_{n,i_1,j_1})\geq \ell)\geq\Prob(\nu(\U_{n,i_2,j_2})\geq \ell).$$
Consequently, Equation \eqref{eq:cij} implies $\Prob(\nu(\M_n)\geq \ell)$ is
bounded from above by
$$\begin{array}{l}
\displaystyle\Prob(\nu(\U_{n,a_n,b_n})\geq\ell)\sum_{\substack{0\leq i\leq
a_n\\0\leq j\leq b_n}} c_{n,i,j}\,\,+\\\sum_{\substack{0\leq i\leq
\binom{n+1}2\\b_n+1\leq j\leq n+1}}\Prob(\nu(\U_{n,i,j})\geq\ell)\,c_{n,i,j}+\\
\sum_{\substack{a_n+1\leq i\leq \binom{n+1}2\\0\leq j\leq b_n}}\Prob(\nu(\U_{n,i,j})\geq\ell)\,c_{n,i,j}.\end{array}$$
Note that the first summand is bounded by $\sum_{\substack{0\leq i\leq
    a_n\\0\leq j\leq b_n}} c_{n,i,j}$, the second summand is
bounded by $\sum_{\substack{0\leq i\leq \binom{n+1}2\\b_n+1\leq j\leq
    n+1}}c_{n,i,j}$ and the third one is bounded by $\sum_{\substack{a_n+1\leq i\leq \binom{n+1}2\\0\leq
    j\leq b_n}}c_{n,i,j}$. Since the sum of these bounds equals
$1$, and since the left-hand side of the inequality tends to $1$ as
$n$ grows by Lemma \ref{lem:mainProbaResult}, if
$\underset{n\to\infty}{\liminf} \sum_{\substack{0\leq i\leq a_n\\0\leq
    j\leq b_n}} c_{n,i,j}>0$, then
$\Prob(\nu(\U_{n,a_n,b_n})\geq\ell)$ must tend to $1$ as $n$ grows.

We prove now that $\underset{n\to\infty}{\liminf}
\sum_{\substack{0\leq i\leq a_n\\0\leq j\leq b_n}} c_{n,i,j}\geq 1/4$. First,
we rewrite $\sum_{\substack{0\leq i\leq a_n\\0\leq j\leq b_n}} c_{n,i,j}$ as
\begin{equation}\label{eq:productbinom}\left(\sum_{0\leq i\leq a_n} p_n^i(1-p_n)^{\binom{n+1}2-i}\binom{\binom{n+1}2}i\right)\left(\sum_{0\leq j\leq b_n} q_n^j(1-q_n)^{n+1-j}\binom{n+1}j\right).\end{equation}
Notice that if $(\mathcal T_n)$ is a sequence of random variables following a binomial distribution $B(n,s_n)$ (\emph{i.e.} the sum of $n$ Bernoulli independent variables of parameter $s_n$) such that $s_n\underset{n\to\infty}{\rightarrow}0$ and $n s_n\underset{n\to\infty}{\rightarrow}\infty$, then $(\mathcal T_n-ns_n)/\sqrt{n s_n}$ converges in distribution to the standard Gaussian distribution $\mathcal N$ (this can be seen on the pointwise convergence of the moment generating function). This implies 
$$\liminf_{n\to\infty}\Prob(\mathcal T_n\leq n s_n)\geq \Prob(\mathcal N\leq 0)=1/2,$$
where $\mathcal N$ is a standard Gaussian distribution. Then, we remark that by construction the first factor in Eq. \eqref{eq:productbinom} equals $\Prob(\lvert\MNS_n\rvert\leq \binom{n+1}{2}p_n)$ and $\lvert\MNS_n\rvert$ follows a binomial distribution of parameters $(\binom{n+1}2,\Omega(1/n))$. Therefore, we obtain 
$$\underset{n\to\infty}{\liminf}\left(\sum_{0\leq i\leq a_n} p_n^i(1-p_n)^{\binom{n+1}2-i}\binom{\binom{n+1}2}i\right)\geq 1/2.$$ A similar argument shows the same lower bound for the second factor in Eq. \eqref{eq:productbinom}, finishing to prove that
$\underset{n\to\infty}{\liminf}
\sum_{\substack{0\leq i\leq a_n\\0\leq j\leq b_n}} c_{n,i,j}\geq 1/4>0$. 

As explained above, this implies that $\underset{n\to\infty}{\lim}\Prob(\nu(\U_{n,a_n,b_n})\geq\ell)=1$ for any $\ell\in\N$.
Finally,
$$\begin{array}{r@{~}c@{~}l}
\Prob(\nu(\U_{n,a_n,b_n})\geq\ell)&=&\Prob(\nu(\U_{n,a_n,b_n})\geq (k^2+3k+2)/2)\\
&=&\displaystyle\Prob\left(\frac{\sqrt{1+8\nu(\U_{n,a_n,b_n})}-1}{2}\geq
k+1\right)\\&\underset{n\rightarrow\infty}{\longrightarrow}&1.
\end{array}$$
This proof is concluded by noticing that $k+1=\lvert \U_{n,a_n,b_n}\rvert-n$.

\end{document}